\documentclass[aip,reprint,floatfix]{revtex4-1}      

\draft{}    

\usepackage{graphicx}       
\graphicspath{{figures/}}   

\usepackage{amsfonts}       
\usepackage{mathtools}
\usepackage{dsfont}
\usepackage{stackrel}

\DeclarePairedDelimiter\floor{\lfloor}{\rfloor}
\DeclareMathOperator*{\argmax}{arg\,max}

\usepackage{glossaries}     
\newacronym{DFT}{DFT}{Density Functional Theory}
\newacronym{ML}{ML}{Machine Learning}
\newacronym{MMS}{MMS}{Myopic Multiscale Sampling}
\newacronym{LSE}{LSE}{Level Set Estimation}
\newacronym{CFGP}{CFGP}{Convolution-Fed Gaussian Process}
\newacronym{GASpy}{GASpy}{the Generalized Adsorption Simulator for Python}
\newacronym{eV}{eV}{electron volts}
\newacronym{CDF}{CDF}{cumulative density function}
\newacronym{dF}{$\Delta \hat{F}$}{predicted change in $F1$ score}

\usepackage{color}

\begin{document}


\title{Computational catalyst discovery:  Active classification through myopic multiscale sampling}

\author{Kevin Tran}
\altaffiliation{These authors contributed equally to this work}
\affiliation{Chemical Engineering Department, Carnegie Mellon University, Pittsburgh, PA 15217}
\author{Willie Neiswanger}
\altaffiliation{These authors contributed equally to this work}
\affiliation{Machine Learning Department, Carnegie Mellon University, Pittsburgh, PA 15217}
\author{Kirby Broderick}
\affiliation{Chemical Engineering Department, Carnegie Mellon University, Pittsburgh, PA 15217}
\author{Eric Xing}
\affiliation{Machine Learning Department, Carnegie Mellon University, Pittsburgh, PA 15217}
\author{Jeff Schneider}
\affiliation{Machine Learning Department, Carnegie Mellon University, Pittsburgh, PA 15217}
\author{Zachary W. Ulissi}
\email{zulissi@andrew.cmu.edu}
\affiliation{Chemical Engineering Department, Carnegie Mellon University, Pittsburgh, PA 15217}

\date{\today}

\begin{abstract}
    The recent boom in computational chemistry has enabled several projects aimed at discovering useful materials or catalysts.
    We acknowledge and address two recurring issues in the field of computational catalyst discovery.
    First, calculating macro-scale catalyst properties is not straight-forward when using ensembles of atomic-scale calculations (e.g., density functional theory).
    We attempt to address this issue by creating a multi-scale model that estimates bulk catalyst activity using adsorption energy predictions from both density functional theory and machine learning models.
    The second issue is that many catalyst discovery efforts seek to optimize catalyst properties, but optimization is an inherently exploitative objective that is in tension with the explorative nature of early-stage discovery projects.
    In other words:  why invest so much time finding a ``best'' catalyst when it is likely to fail for some other, unforeseen problem?
    We address this issue by relaxing the catalyst discovery goal into a classification problem:  ``What is the set of catalysts that is worth testing experimentally?''
    Here we present a catalyst discovery method called myopic multiscale sampling, which combines multiscale modeling with automated selection of density functional theory calculations.
    It is an active classification strategy that seeks to classify catalysts as ``worth investigating'' or ``not worth investigating'' experimentally.
    Our results show a $\sim$7--16 times speedup in catalyst classification relative to random sampling.
    These results were based on offline simulations of our algorithm on two different datasets:  a larger, synthesized dataset and a smaller, real dataset.
\end{abstract}


\maketitle


\section{Introduction}

Recent advances in computing hardware and software have led to substantial growth in the field of computational materials science.
In particular, databases of high-throughput calculations\cite{Jain2013, Curtarolo2013, Saal2013, Ong2013, Berman2000, Meyer2010} have increased the amount of information available to researchers.
These databases facilitate the development of models that supplement human understanding of physical trends in materials.\cite{SchlexerLamoureux2019, Schleder2019, Medford2018}
These models can then be used in experimental discovery efforts by identifying promising subsets of the search space, resulting in increased experimental efficiency.\cite{Chung2014, Ioannidis2016, Chakraborty2017, Green2017, Tran2018, Zhong2020}

However, many materials design efforts use material properties and calculation archetypes that are too problem-specific to be tabulated in generalized databases.
When such efforts coincide with design spaces too large to search in a feasible amount of time, we need a way to search through the design space efficiently.
Sequential learning, sometimes referred to as optimal design of experiments or active learning, can fill this role.
Sequential learning is the process of using the currently available data to decide which new data would be most valuable for achieving a particular goal.\cite{Lookman2019, Schmidt2019, Kim2019a}
In practice, this usually involves fitting a surrogate model to the available data and then pairing the model with an \textit{acquisition function} that calculates the values of a new, potential data points.
Then we \textit{query} the most valuable data points, add them to the data set, and repeat this process.
These sequential learning methods have been estimated to accelerate materials discovery efforts by up to a factor of 20.\cite{Rohr2020}

Sequential learning has numerous sub-types of methods that can and have been used for different goals.
One such sub-type is active learning.
With many active learning algorithms, the goal is to replace a relatively slow data-querying process with a faster-running surrogate model.\cite{Settles2012}
Since the surrogate model may be used to query any point, the acquisition functions focus on ensuring that the entire search space is explored.
Another sub-type of sequential learning is active optimization.\cite{Frazier2018}
With this sub-type, the goal is to maximize or minimize some objective function.
Thus the acquisition functions generally focus on parts of the search space where maxima or minima are more likely to occur.
One of the most common types of active optimization is Bayesian optimization.\cite{Frazier2018}
Yet another sub-type of sequential learning is online or on-the-fly learning.\cite{Hoi2018}
The goal for these methods is to accelerate the predictions of streams of data.
In the field of computational material science, this is often applied to predicting trajectories for \gls{DFT} or molecular dynamics calculations.\cite{Khorshidi2016, Vandermause2020}

In computational materials discovery, we often have the following task:  we have a set of available materials $\mathcal{X} = \{x_i\}_{i=1}^n$, where each material $x_i$ has an associated quantity $y_i$, denoting its value for some application.  
Examples of common properties for $y_i$ include---but are not limited to---formation energies of materials, catalyst activity, tensile strength, or conductivity.
The value $y_i$ is unknown and must be calculated, which can be costly in time, money, or other resources.
Further, theoretical calculations of material properties may be inconsistent with experimental results.
As per a common aphorism among statisticians:  ``All models are wrong, but some are useful.''

Due to these potential model errors and due to the exploratory nature of materials discovery, we propose reframing the materials discovery question.
Instead of trying to discover materials with optimal $y_i$ values, what if we instead classify materials as having promising or unpromising $y_i$ values?
In other words, what if we frame materials discovery efforts as classification problems rather than optimization problems?
The estimated classes could then be used to design physical experiments.
Mathematically, this is akin to assuming that material $i$ has a binary value $y_i \in \{0, 1\}$, where $0$ denotes \textit{``not of interest''}, and $1$ denotes \textit{``of interest''}.

The goal is then to determine the values $y_i$ for each $x_i \in \mathcal{X}$ as cheaply as possible.
One can view this as the task of most-efficiently learning a classifier that, for each $x_i$, correctly predicts its value $y_i$.
In this way, materials discovery problems can be framed as problems of \textit{active classification}.
Active classification is the task of choosing an ordering of $x_i \in \mathcal{X}$, over which we will iterate and sequentially measure their values $y_i$, in order to most efficiently (using the fewest measurements) learn a classifier that predicts the correct label for all materials $x_i \in \mathcal{X}$.\cite{Ma2015, Zanette2019}

Another aspect of computational materials discovery is the ability to turn calculations into recommendations---e.g., how can we convert \gls{DFT} results into actionable experiments?
This conversion is relatively straight-forward when properties are directly calculable, which is the case for properties such as the enthalpy of formation.\cite{Flores2020}
If we perform a single \gls{DFT} calculation that suggests a single material may be stable, then we can suggest that single material for experimentation.
But for many applications, the properties of interest may not be calculable directly.
For example, let us say we are interested in finding active catalysts.
One way to do this is to use \gls{DFT} to calculate the adsorption energy between the catalyst and particular reaction intermediates, and then couple the resulting adsorption energy with a Sabatier relationship.\cite{Seh2017}
But \textit{in situ}, a catalyst comprises numerous adsorption sites and surfaces.
Thus the true activity of a catalyst may be governed by an ensemble of adsorption energies, and therefore may need multiple \gls{DFT} calculations.
How do we address the fact that we need multiple \gls{DFT} queries to resolve the properties of a single material?

\begin{figure*}
    \vspace{5mm}
    \centering
    \includegraphics[width=\textwidth]{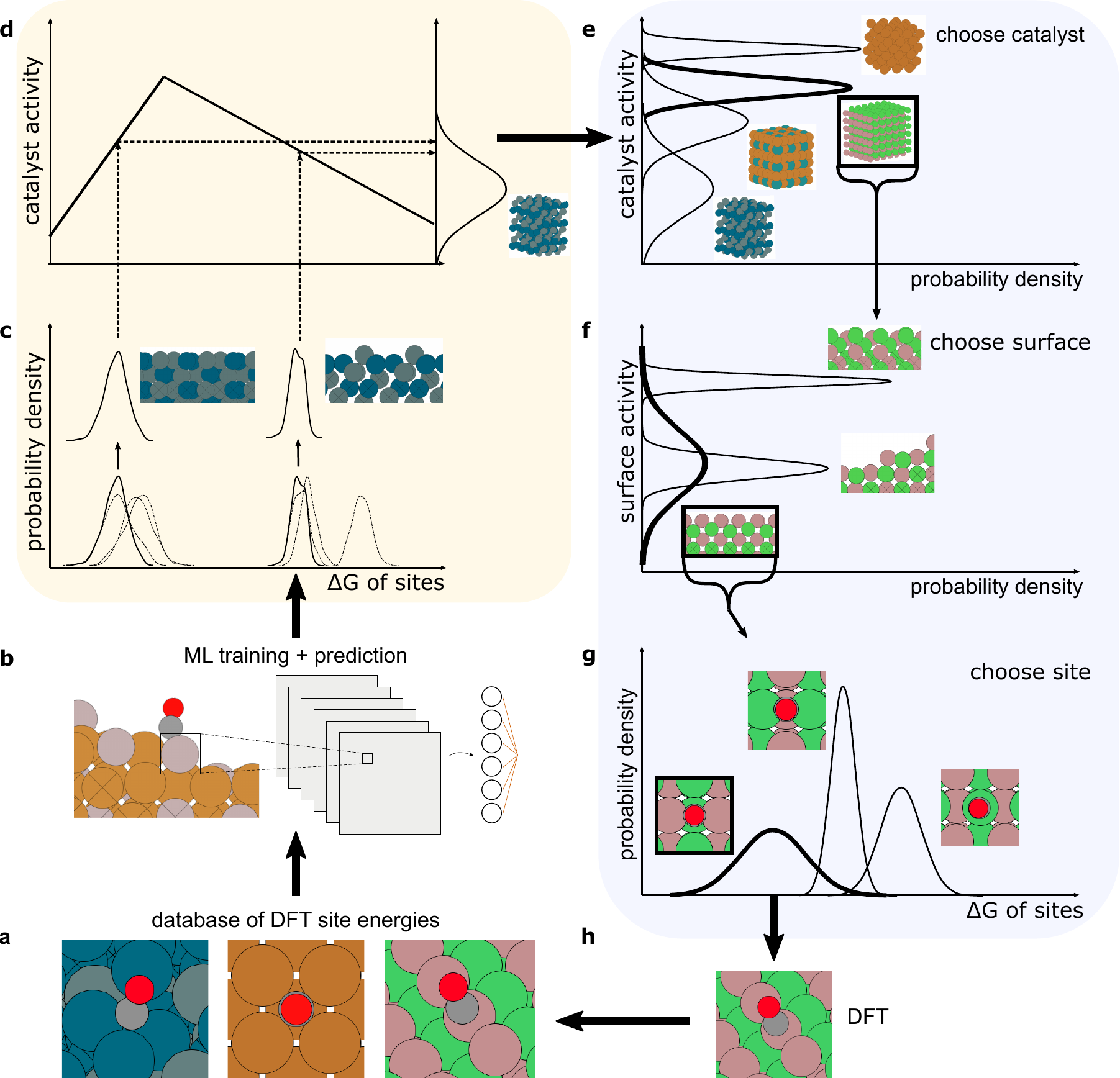}
    \vspace{5mm}
    \caption{Illustration of \glsentryfull{MMS}.
        Given a database of \gls{DFT}-calculated adsorption energies (\textbf{a}), we train a \gls{ML} model to predict adsorption energies (\textbf{b}).
        Then we use those adsorption energies to estimate activities of catalyst surfaces (\textbf{c}), which we then use to estimate the activities of the bulk catalysts (\textbf{d}).
        Then we choose which catalyst to sample next (\textbf{e}); then we choose which surface on the catalyst to sample (\textbf{f}); then we choose which site on the surface to sample (\textbf{g}); then we perform \gls{DFT} of that site to add to the database (\textbf{h}).
        This procedure is repeated continuously with the goal of classifying all catalysts as either ``relatively active'' or ``relatively inactive''.}\label{fig:overview}
\end{figure*}

Here we attempt to address both outlined issues:  (1) we need an ensemble of \gls{DFT} queries to calculate a single experimental property of interest, and (2) we need a sequential learning method designed for high-throughput discovery/classification.
We overcome both issues by creating the \gls{MMS} method (Figure~\ref{fig:overview}).
\gls{MMS} addresses the first aforementioned issue by using a multiscale modeling framework for estimating the activity of a catalyst using an ensemble of both \gls{DFT} and \gls{ML} predicted adsorption energies.
\gls{MMS} then addresses the second issue by combining this multiscale modeling framework with a number of sequential learning methods, including active classification.
Note that \gls{MMS}, as we describe it in this paper, is tailored to discovering active catalysts.
Although this method may not be directly transferable to other applications, we hope that others may be able to adapt the principles of the method to their own applications.

\section{Methods}

\subsection{Multiscale Modeling}

In this paper, we use the discovery of active catalysts as a case study.
Catalyst activity is often correlated with the adsorption energy of particular reaction intermediates, as per the volcano relationships stemming from the Sabatier principle.\cite{NorskovBook, Seh2017}
These adsorption energies can be calculated using \gls{DFT}.
Each \gls{DFT}-calculated adsorption energy is specific to a particular binding site of a particular surface of a particular catalyst.
Thus the relationship between \gls{DFT}-calculated adsorption energies and a catalyst's activity is not simple.

For example:  in cases of lower adsorbate coverage on the catalyst surface, adsorbates tend to adsorb to stronger-binding sites before weaker-binding sites.
In cases of higher adsorbate coverage, adsorption energies are difficult to calculate, so it is not uncommon to assume low adsorbate coverage.\cite{NorskovBook,Norskov2005, Lopato2020}
It follows that the activity of a surface could be estimated by using the Sabatier-calculated activity of the strongest binding site on a surface.

Given the activities of the surfaces of a catalyst, the next step is to estimate the activity of the entire catalyst.
One way to do this would be to perform a weighted average of the surface activities, where higher weights are given to surfaces that are more stable.
For simplicity's sake, we instead propose a uniform average and recognize that future work may involve investigating more sophisticated averaging methods.

\begin{figure}
    \centering
    \includegraphics[width=0.45\textwidth]{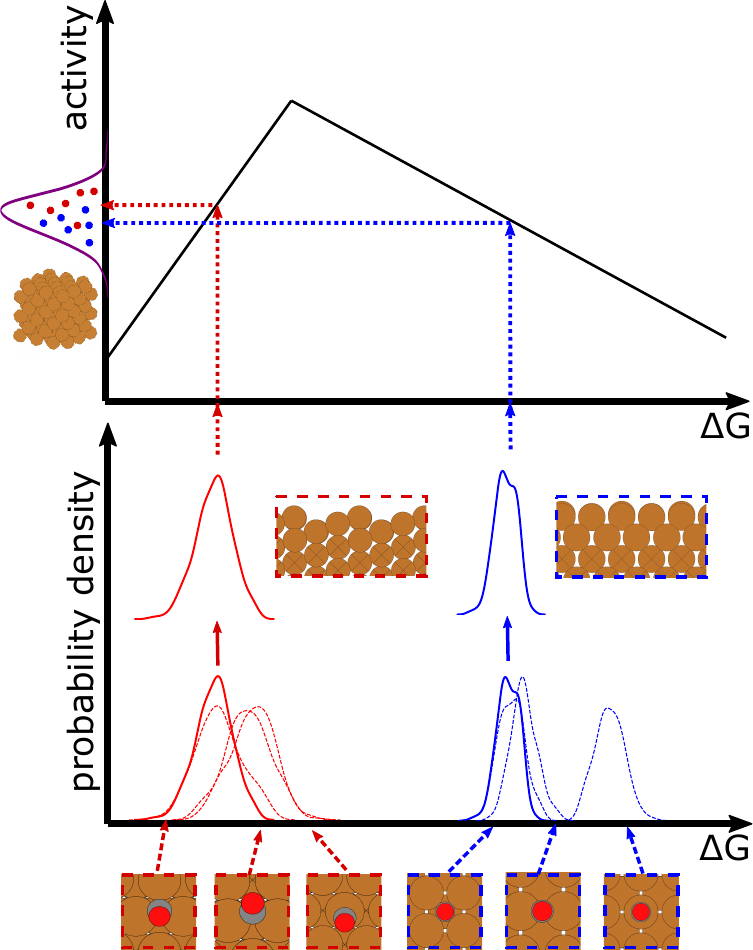}
    \caption{Multiscale modeling strategy for estimating the activity of a catalyst.
        For each adsorption site, we obtain a machine-learned estimate of its adsorption energy along with uncertainty.
        Then we aggregate the energy distributions for all sites within each surface through a linear function of sites.
        Next we transform the energy distributions for all surfaces into activities using a Sabatier relationship.
        Finally we average all the surface activities to obtain an estimate of overall catalyst activity.}\label{fig:modeling}
\end{figure}

Concretely, suppose we have $n$ catalyst candidates $\{x_i\}_{i=1}^n$, where each candidate $x_i$ has $m$ surfaces $\{u_{i, j}\}_{j=1}^m$, and surface $u_{i, j}$ has $\ell$ sites $\{s_{i,j,k}\}_{k=1}^\ell$.  
For a given site $s_{i,j,k}$, denote its adsorption energy by $\Delta G(s_{i,j,k})$, and for a given surface $u_{i,j}$, denote its catalytic activity by $\alpha(u_{i,j})$.
Likewise, for a given catalyst material candidate $x_i$, denote the average catalytic activity for the candidate by $\alpha(x_i) = \frac{1}{m} \sum_{j=1}^m \alpha(u_{i, j})$.
Suppose we have a predictive uncertainty estimate for the adsorption energy $\Delta G(s_{i,j,k})$ of a site, represented by a Normal distribution with mean $\mu_{i,j,k}$ and variance $\sigma^2_{i,j,k}$.
We can then perform simulation-based uncertainty quantification of catalyst activity by using the multiscale modeling process we described above to propagate uncertainties from sites' adsorption energies.
Specifically, for each material candidate $x_i$, we generate $H$ samples of its catalytic activity, $\{\tilde{\alpha}_i^h\}_{h=1}^H$, by simulating from the following generative process:  

\begin{align}
    \label{eq:genprocess}
    \text{For } j&=1,\ldots,m, \hspace{1mm} k = 1,\ldots,\ell:\\
        &\{\tilde{\Delta}G_{i,j,k}^h\}_{h=1}^H \stackrel{iid}{\sim}  
            \mathcal{N}\left(\mu_{i,j,k}, \sigma^2_{i,j,k}\right) \nonumber\\
    \text{For } h&=1,\ldots,H, \hspace{1mm} j=1,\ldots,m: \nonumber\\
    \tilde{\alpha}_{i,j}^h &= 
    \begin{cases}
        \exp(M_1 \tilde{\Delta}G_{i,j,1:\ell}^h + B_1) & \text{if } \tilde{\Delta}G_{i,j,1:\ell}^h \geq t^* \\
        \exp(M_2 \tilde{\Delta}G_{i,j,1:\ell}^h + B_2) & \text{otherwise}
    \end{cases} \nonumber \\
    \text{For } h&=1,\ldots,H: \nonumber\\
        &\tilde{\alpha}_i^h = \frac{1}{n} \sum_{j=1}^m \tilde{\alpha}_{i,j}^h \nonumber
\end{align}

\noindent
where $t^*$ is the optimal absorption energy for a given volcano relationship and $M_1$, $M_2$, $B_1$, \& $B_2$ are the linear coefficients associated with the two sides of the log-scaled volcano relationship of a given chemistry.
Figure~\ref{fig:modeling} illustrates how we use our multiscale modeling method to estimate catalyst activity from \gls{DFT}-calculated adsorption energies, including uncertainty quantification.

Each catalyst material candidate $x \in \mathcal{X}$ has some true catalytic activity level $\alpha(x)$.
Our goal will be to determine the top $p$-\% of catalyst material candidates in terms of their activity levels, which we denote $\mathcal{X}_p = \{x \in \mathcal{X} : r(\alpha(x)) \geq \floor*{ \frac{pn}{100} } \}$, where $r: \mathbb{R}_+ \rightarrow \{1, \ldots, n\}$ is a function mapping the activity level $\alpha(x)$ to an index denoting it's rank (from highest to lowest activity).
Given a specified $p$, if a candidate material is in this set, i.e. $x_i \in \mathcal{X}_p$, then we say that its associated binary value $y_i=1$, and say $y_i=0$ otherwise.
In simpler terms:  we want to find the top $p$-\% most active catalysts.
For this paper, we choose $p = 10\%$ arbitrarily.
Any catalyst that falls within the top 10\% in terms of activity will be labeled as active, and anything below the top 10\% will be labeled as inactive.

We can therefore frame our goal as determining the associated binary value $y_i$ for each catalyst material candidate $x_i \in \mathcal{X} = \{x_i\}_{i=1}^n$.  
Suppose we have formed point estimates for each of the binary values, written $\{\hat{y}_i\}_{i=1}^n$.  
To assess the quality of this set of estimates with respect to the set of true candidate values, we focus on the $F_1$ score---a popular metric for classification accuracy, defined as

\begin{align}
    F_1 &= 2 \times \frac{\text{precision} \times \text{recall}}{\text{precision} +
        \text{recall}}\\
    &= \frac{2 \sum_{i=1}^n y_i \hat{y}_i}{2 \sum_{i=1}^n y_i \hat{y}_i +
        \sum_{i=1}^n (1 - y_i) \hat{y}_i + \sum_{i=1}^n y_i (1 - \hat{y}_i)}. \nonumber
\end{align}

\noindent
Given a set of ground-truth values $\{y_i\}_{i=1}^n$, we are able to compute the $F_1$ score for a chosen set of value estimates $\{\hat{y}_i\}_{i=1}^n$.  

However, in practice, we will typically not have access to these ground-truth values, and thus cannot compute this score in an online procedure.
For use in online experiments, we will take advantage of a metric that yields an
estimate of the change in $F_1$ score. This metric is computable using only our model
of the activity of each catalyst, without requiring access to ground-truth values
$\{y_i\}_{i=1}^n$, and can be used to assess and compare the convergence of our methods.
Furthermore, it can be used to provide an early stopping method for our active procedures.
We will show experimentally in Section~\ref{sec:results} that this metric shows a strong correlation to the $F_1$ score.

\subsection{Sampling Strategy}

\begin{figure}
    \centering
    \includegraphics[width=0.45\textwidth]{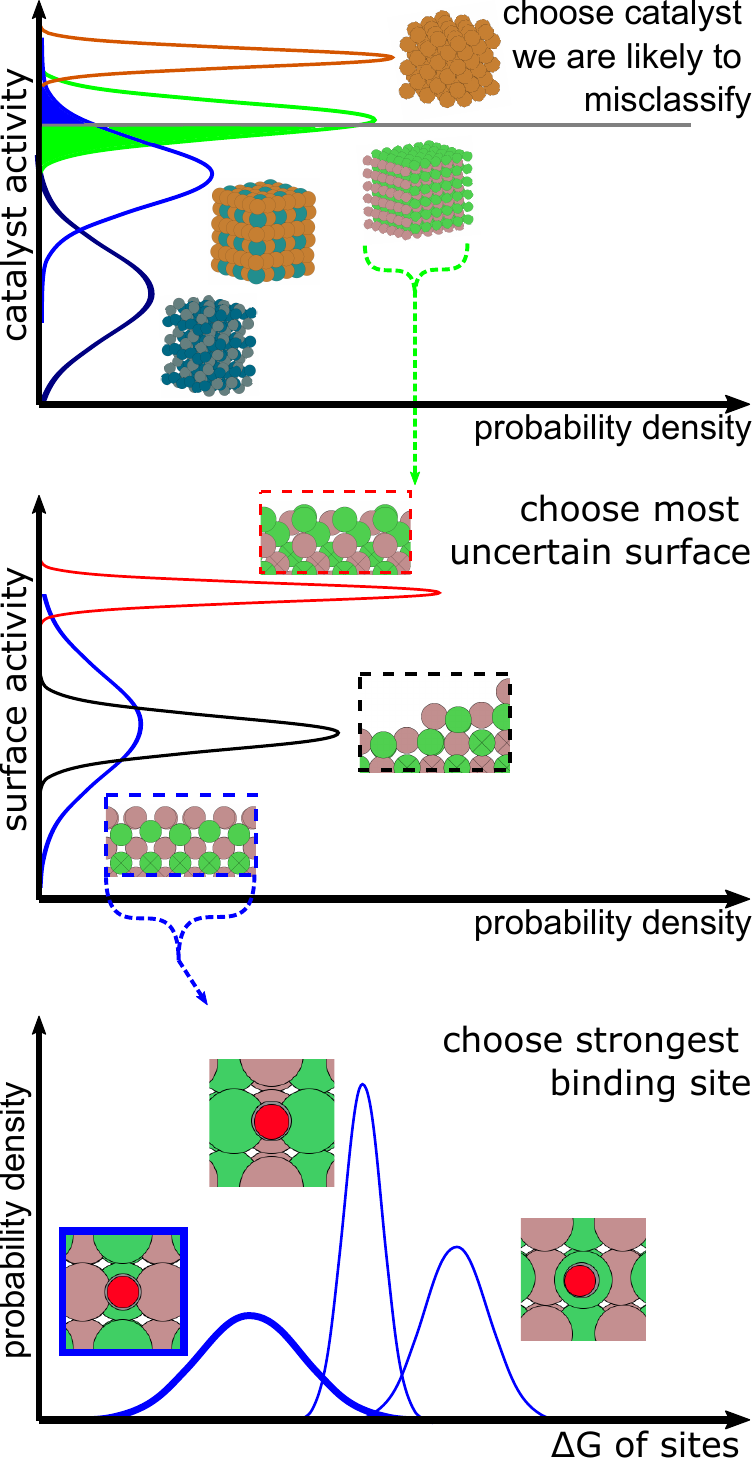}
    \caption{\glsentryfull{MMS} overview.
        At the highest level, we choose a catalyst to query using level-set estimation---specifically, we use the probability of incorrect classification as our acquisition function.
        At the middle level, we choose a surface of the catalyst using uncertainty sampling.
        At the lowest level, we choose a site on the surface using Bayesian optimization to find the lowest energy site.}\label{fig:mms}
\end{figure}

The goal of \gls{MMS} is to discover catalysts that are likely to be experimentally active.
Optimization of catalytic activity is not the main priority, because we assume that unforeseen experimental issues are likely to obsolete most candidate catalysts.
Instead, a greater focus is given on identification of a large number of candidates rather than finding ``the most active'' candidate.
That is why the core sequential learning algorithm we use in \gls{MMS} is active classification.\cite{Ma2015, Zanette2019}
To be specific, we use \gls{LSE} to identify catalysts for \gls{DFT} sampling.
After identifying catalysts for \gls{DFT} sampling, we then need to choose which surface of the catalyst to sample; here we use techniques from active regression.
Once a surface is chosen, we then attempt to find the strongest binding site on that surface by using active optimization of the adsorption energies.
Thus we combine three different sequential learning strategies across three different length scales to decide which site-based \gls{DFT} calculation will help us classify active vs.\ inactive catalysts (Figure~\ref{fig:mms}).

We first describe the initial step of our sampling strategy, which consists of selecting a catalyst material candidate from our candidate set $\mathcal{X} = \{x_i\}_{i=1}^n$.  
Note that our high-level goal is binary classification, in that we want to efficiently produce accurate estimates $\{\hat{y}_i\}_{i=1}^n$ of the binary value for each material candidate.  
Based on our definition of $y_i = \mathds{1}\left[x_i \in \mathcal{X}_p \right]$, this problem can be equivalently viewed as the task of \gls{LSE}, in which we aim to efficiently produce an accurate estimate of the superlevel set $\mathcal{X}_p = \{x \in \mathcal{X} : r(\alpha(x)) \geq \floor*{ \frac{pn}{100} } \}$.
There has been a body of work on developing acquisition functions for choosing candidates to query in the task of \gls{LSE}.\cite{gotovos2013active,kandasamy2019myopic}
In particular, we focus on the \textit{probability of incorrect classification} acquisition function,\cite{bryan2005active} defined for an $x_i \in \mathcal{X}$ as

\noindent
\begin{align}
    \varphi(x_i) &= \min (p, 1-p), \text{where}\\
    p &= \text{Pr}\left( r(\alpha(x)) \geq \floor*{\frac{pn}{100}} \right) \nonumber \\
    &\approx
    \underbrace{ \frac{1}{H} \sum_{h=1}^H
    \mathds{1}\left[r(\tilde{\alpha}_i^h) \geq
    \floor*{\frac{pn}{100}}\right]}_\text{Empirical probability $\alpha(x)$ in top $p$-\%} \nonumber
\end{align}

\noindent
Thus to select a subsequent catalyst candidate, we compute $\varphi(x_i)$ for each $x_i \in \mathcal{X}$ and return the maximizer $x^* = \argmax_{x_i \in \mathcal{X}} \varphi(x_i)$.
In simpler terms:  we choose the catalyst that we are most likely to classify incorrectly.
Note how this implies that we \textit{not} query catalysts that we are confident are active, which is different from active optimization methods.
This provides a more exploratory method rather than an exploitative one, which is appropriate in early-stage computational discoveries and screenings.

The selection of a catalyst candidate $x_i$ depends on its estimated catalytic activity, which we model as an average of the catalytic activities across the surfaces of the candidate, i.e. $\alpha(x_i) = \frac{1}{m} \sum_{j=1}^m \alpha(u_{i, j})$.
Though we select a candidate based on its ability to help improve our estimate of the superlevel set $\mathcal{X}_p$, once selected, we then wish to most efficiently improve our estimate of this candidate's catalytic activity.
Our goal at this stage is therefore to most efficiently learn the catalytic activities for each surface of that candidate.
This can be viewed as an active regression task, where we aim to sample a surface that will most reduce the uncertainty of our surface activity estimates.
To select a surface, we use an \textit{uncertainty sampling for regression} acquisition function from the active learning literature\cite{settles2009active}, defined as

\begin{align}
    \varphi(u_{i, j}) &=
    \text{Var} \left[ \text{Pr}\left( \alpha(u_{i, j}) \right) \right]\\
    &\approx \frac{1}{H-1} \sum_{h=1}^H
    \left( \tilde{\alpha}_{i, j}^h -
        \frac{1}{H} \sum_{h'=1}^H \tilde{\alpha}_{i,j}^{h'} \right)^2, \nonumber  
\end{align}

\noindent
which selects a surface $u_i^*$ of material candidate $x_i$ that has the greatest variance.
In simpler terms:  we choose the surface of a catalyst that has the most uncertainty, because we suspect that this choice is most likely to reduce our uncertainty estimate of catalyst activity.

The catalytic activity of a given surface $\alpha(u_{i,j})$ is function of the adsorption energies of the sites on this surface, according to the relationship
$\alpha(u_{i,j}) =  \exp(- | M \tilde{\Delta}G_{i,j,1:\ell} + B |)$  from Equation (\ref{eq:genprocess}),
where $\tilde{\Delta}G_{i,j,1:\ell}$ is the set of adsorption energies over all sites on the surface.
Therefore, given a selected surface $u_{i,j}$, we wish to determine efficiently the site on this surface with minimum adsorption energy.
This can be viewed as an optimization task.
We therefore use the \textit{expected improvement} acquisition function from Bayesian optimization\cite{movckus1975bayesian}, defined as

\begin{align}
    \varphi(s_{i, j, k}) &=
    \mathbb{E}\left[
        (\Delta G(s_{i,j,k}) \leq \Delta G^*) \mathds{1}\left[ \Delta G(s_{i,j,k}) - \Delta G^*\right]
    \right] \nonumber \\
    &\approx
        \Phi\left(\frac{\Delta G^* - \tilde{\mu}_{i,j,k}}{\tilde{\sigma}_{i,j,k}}\right)
        \phi \left(\frac{\Delta G^* - \tilde{\mu}_{i,j,k}}{\tilde{\sigma}_{i,j,k}}\right) \\
    & \hspace{4mm} \times \left(\Delta G^* - \tilde{\mu}_{i,j,k} \right), \nonumber
\end{align}

\noindent
where $\tilde{\mu}$ $=$ $\frac{1}{H}\sum_{h=1}^H \tilde{\Delta G}_{i,j,k}^h$ is the
expected adsorption energy, $\tilde{\sigma}$ $=$ $\sqrt{\frac{1}{H-1}\sum_{h=1}^H
\left(\tilde{\Delta G}_{i,j,k}^h - \tilde{\mu} \right)^2}$ is its standard deviation,
$\Phi$ is the \gls{CDF} of a standard normal distribution, $\phi$ is the PDF of a
standard normal distribution, and $\Delta G^*$ is the minimum observed adsorption
energy.  This selects a site $s_{i,j}^*$ which is expected to most reduce the site
adsorption energy relative to the current minimum observed energy, and allows for
efficient estimation of the minimum energy site on surface $u_{i,j}$.  In simpler terms:
we choose the site on a surface that is most likely to help us identify the
strongest/lowest binding site on the surface.

\subsection{Active Learning Stopping Criteria}

Assessing convergence of an active algorithm is useful for enabling early stopping, which can save resources.
Measures of convergence can also provide diagnostics in online use settings.
To quantify convergence, we use the \textit{\gls{dF}}\cite{Altschuler2019}.
Intuitively speaking, this rule says to stop an active learning procedure when \gls{dF} drops below a predefined threshold $\epsilon$ when for $k$ consecutive windows, i.e.,

\begin{align}
\textit{Stop} \hspace{3mm} &\text{if } \Delta\hat{F} < \epsilon \text{ over } k \text{ windows} \nonumber \\
\textit{Continue} \hspace{3mm} &\text{otherwise}. \nonumber
\end{align}
In our setting, $\Delta\hat{F}$ is defined to be
\begin{align}
    \hat{\Delta}F = 1 - \frac{2a}{2a + b + c},
\end{align}

\noindent
where $a$ is the number of bulks for which the model at iterations $i$ and $i+1$ both yield a positive label, $b$ is the number of bulks for which the model at iteration $i$ yields a positive label while at iteration $i+1$ yields a negative label, and $c$ is the number of bulks for which the model at iteration $i$  yields a negative label while at iteration $i+1$ yields a positive label.
Each of $a$, $b$, and $c$ are computed over the previous $k$ iterations.
This measure provides an estimate of the change in accuracy at each iteration, and it allows us to control how conservatively (or aggressively) we stop early via an interpretable parameter $\epsilon$.
We show results of this measure alongside our $F1$ score in Section~\ref{sec:results}.
Note that Altschuler \& Bloodgood\cite{Altschuler2019} recommend using a \textit{stop set} of unlabeled points over which to calculate \gls{dF}.
Here we use the entire search space of catalysts in lieu of a stop set, because it was non-trivial for us to define a stop set that was representative of the search space.

\subsection{Management of Data Queries}

Implementation of \gls{MMS} also involves definition of several hyper-parameters.
For example, most surrogate models require training data before making predictions to feed the sampling method.
This means that we needed to seed \gls{MMS} with initial training data.
We chose to create the initial training data by randomly sampling 1,000 adsorption energies from the search space.
We used random sampling for simplicity, and we sampled 1,000 adsorption energies because that was the minimum amount of data on which \gls{CFGP} (described below in further detail) could train on and maintain numerical stability.

Another consideration for \gls{MMS} is the \textit{batch size} and how to handle queries in-tandem.
Normal sequential learning assumes that we can make one query at a time.
But in applications such as ours, it may be possible to make multiple queries in parallel---i.e., we can perform multiple \gls{DFT} calculations at a time.
There are several methods for handling queries in parallel; we chose to use a type of look-ahead sampling.\cite{Desautels2012}
With look-ahead sampling, we began by choosing the first point to sample using the standard acquisition strategy.
Then, while that point was still ``being queried'', we assumed that the first point was queried successfully and set the ``observed'' value equal to our predicted value.
In other words, we pretend that we sampled the first data point and that our prediction of it was perfect.
This allowed us to then recalculate our acquisition values to choose a second point.
This process of ``looking ahead'' one point at a time was continued until a predetermined number of points were selected for querying---i.e., the batch size.
Here we chose a batch size of 200 points, because that was roughly the number of \gls{DFT} calculations that we could perform in a day during our previous high-throughput \gls{DFT} studies.\cite{Tran2018}
Note that we did not re-train the surrogate models within each batch of 200 points; we only re-calculated acquisition values between each sample within each batch.
We skipped re-training of surrogate models within each batch to reduce the amount of model training time required to perform this study.
Although this may have reduced the effectiveness of the look-ahead method, we found the increased algorithm speed to be worthwhile.

\subsection{Estimating Performance through Simulation}

We aim to experimentally assess the performance of \gls{MMS} and compare it with a variety of baseline methods without incurring the high cost of repeated \gls{DFT} calculations.
To do this, we simulate each procedure using a database of pre-determined adsorption energies.
Specifically, suppose we have chosen a set of $n$ catalyst material candidates $\{x_i\}_{i=1}^n$ of interest.  
For each candidate $x_i$, we already have all the adsorption energies $\Delta G(s_{i,j,k})$ for the full set of sites across the full set of surfaces on $x_i$.
We can then run our procedures in a relatively fast manner, where we can quickly query the database at each iteration of a given method rather than running \gls{DFT}.
Similar offline-data discovery procedures have been pursued by previous work in optimization and active learning, where expensive evaluations have been collected offline and used for rapid online evaluation\cite{Char2020, Ying2019, White2019}.

One notable baseline method is \textit{random search}, which at each iteration samples sites to carry out \gls{DFT} calculations uniformly at random from the full set of sites over all catalyst material candidates.
We provide simulation results using random search as a benchmark to compare \gls{MMS} against.

\subsubsection{Surrogate Models Used}
\label{sec:surrogatemodels}

Our objective in this paper is to assess the performance of \gls{MMS}.
The performance of \gls{MMS} is likely to depend on the surrogate model used to predict adsorption energies from atomic structures.
We assume that surrogate models with high predictive accuracy and calibrated uncertainty estimates\cite{Kuleshov2018} will outperform models with low accuracy and uncalibrated uncertainty estimates, but we are unsure of the magnitude of this difference.
We therefore propose to pair at least two different models with \gls{MMS}:  a ``perfect'' model and an ``ignorant'' model.

We define the ``perfect'' model, hereby referred to as the ``prime'' model, as a model that returns the true adsorption energy of whatever data point is queried.
This perfect prediction ensures a high model accuracy.
When asked for a standard deviation in the prediction, the prime model will return a sample from a $\chi^2$ distribution whose mean is 0.1 \gls{eV}.
This uncertainty ensures a sharp and calibrated\cite{Kuleshov2018, Tran2020} measure of uncertainty.
We do not use standard deviation of zero because (1) it causes numerical issues during multiscale modeling and (2) any model in practice should not be returning standard deviations of zero.

We define the ``ignorant'' model, hereby referred to as the ``null'' model, as a model that returns the optimal adsorption energy no matter what is queried.
This constant prediction ensures a relatively low model accuracy.
When asked for a standard deviation in the prediction, the null model will return 1 \gls{eV}.
This uncertainty ensures a relatively dull and uncalibrated measure of uncertainty.

Lastly, we also choose to use a third, most practical model:  \gls{CFGP}.\cite{Tran2020}
\gls{CFGP} is a Gaussian process regressor whose features are the output of the final convolutional layer in a trained graph convolutional neural network.
This model is our best current estimate of both an accurate and calibrated model that could be used in practice.
Thus we have three models:  null, \gls{CFGP}, and prime, which are intended to give quantitative estimates of the minimal, medial, and maximal performance of \gls{MMS}, respectively.

\subsubsection{Search Spaces Used}

Previous studies have shown that different materials discovery problems have varying difficulties.\cite{Kim2019a}
Searching for a needle in a hay stack is generally more difficult than searching for a leaf on a branch.
Thus any simulation we do depends on the search space we use.
To obtain a range of potential \gls{MMS} performances, we perform simulations using two different data sets.
Both data sets comprise thousands of atomic structures that represent CO adsorbing onto various catalyst surfaces, as well as corresponding adsorption energies.
We then use Sabatier relationships from literature to transform the adsorption energies into estimates of activity.\cite{Liu2017}

We defined our first search space by synthesizing it randomly.
We did so by retrieving a database of enumerated adsorption sites from \gls{GASpy}\cite{Tran2018, Tran2018a}.
These sites composed all the unique sites on all surfaces with Miller indices between -2 and 2 across over 10,000 different bulk crystal structures.
We then randomly selected 200 of the bulk crystals along with all of the resulting surfaces and sites, yielding over 390,000 adsorption sites.
Then for each bulk crystal, we randomly sampled its ``bulk mean adsorption energy'' from a unit normal distribution.
Then for each surface within each crystal, we randomly sampled its ``surface mean adsorption energy'' from a normal distribution whose mean was centered at the corresponding bulk mean and whose standard deviation was set to 0.3 \gls{eV}.
Then for each site within each surface, we randomly sampled its adsorption energy from a normal distribution whose mean was centered at the corresponding surface mean and whose standard deviation was set to 0.1 \gls{eV}.
Thus the adsorption energies were correlated within each bulk, and they were also correlated within each surface.

We defined our second search space by retrieving our database of \textit{ca.} 19,000 \gls{DFT}-calculated CO adsorption energies calculated by \gls{GASpy}, hereafter referred to as the \gls{GASpy} dataset.
The sites in this database were chosen using previous iterations of our sequential learning methods,\cite{Tran2018} and they therefore have bias in the locations at which they were sampled.
Specifically, the sites in this database were chosen based on the likelihood that their adsorption energies were close to the optimal value of -0.67 \gls{eV}.\cite{Liu2017, Tran2018}

There are several advantages of using the synthesized data set over the real \gls{GASpy} data set, and vice versa.
The synthesized data set contains pseudo-random adsorption energies that are difficult for \gls{CFGP} to predict, thereby hindering its performance unfairly.
Therefore, we should not and did not use \gls{CFGP} with the synthesized data set; we used it with the \gls{GASpy} data set only.
On the other hand, the number of surfaces per bulk and the number of sites per surface in the \gls{GASpy} data set was relatively sparse compared to the synthesized data set.
This can result in catalysts that require relatively few site queries to sample fully, which reduces the number of queries necessary to classify a catalyst.
This reduction in the number of required queries per catalyst could artificially improve the observed performance of \gls{MMS}.


\section{Results}
\label{sec:results}

At the beginning of the simulations, the multiscale models made their catalyst class predictions (i.e., active or inactive) using the adsorption energy predictions and uncertainties of the models.
As the simulations progressed and adsorption energies were queried, the models' predictions of each queried energy were replace with the ``true'' value of the query and the corresponding uncertainty was collapsed to 0 \gls{eV}.\
This was done to mimic a realistic use case where we would not use model predictions when we had the ``real'' \gls{DFT} data instead.
It follows that, as the simulations progressed and nearly all points were queried, most models performed similarly because they all had comparable amounts of ``true'' data to use in the multiscale model.

\subsection{Performance on Synthesized Data}

\begin{figure*}
    \centering
    \includegraphics[width=\textwidth]{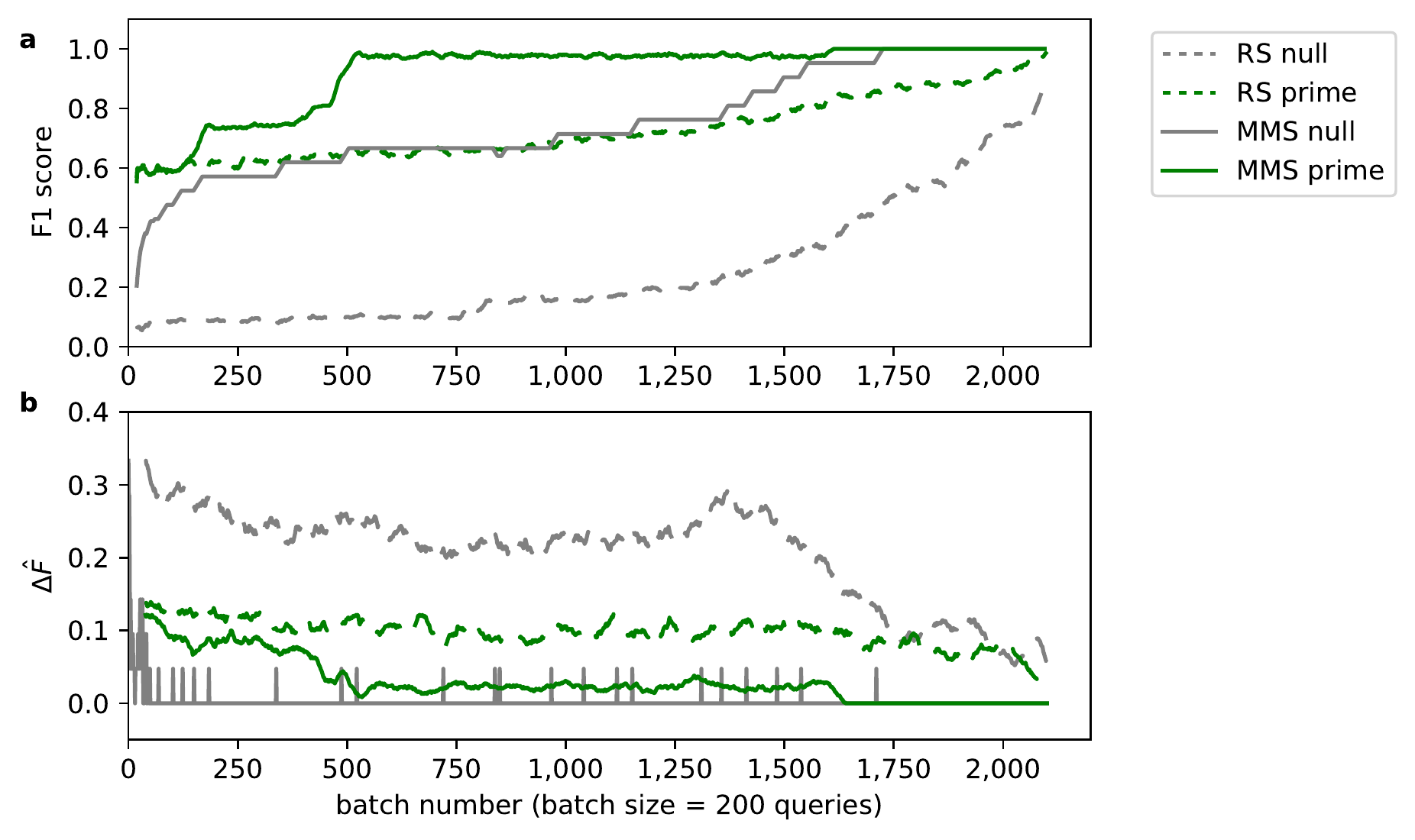}
    \caption{Performance and convergence results for the simulations on the synthesized dataset.
        \textbf{a.} $F1$ score of the multiscale model during simulation of the synthesized data.
        For clarity of visualization, we plotted the rolling average of the $F1$ score using a window of 20 batches.
        \textbf{b.} \gls{dF} of the multiscale model during simulation of the synthesized data.
        For clarity of visualization, we plotted the rolling average of \gls{dF} using a window of 40 batches (excluding the \gls{MMS} null line, where no averaging was done).
        RS represents ``random search'' while \gls{MMS} represents \glsentrylong{MMS}.
        }\label{fig:syn_results}
\end{figure*}

This behavior is seen in Figure~\ref{fig:syn_results}a, which shows how the $F1$ changes at each point in the simulation of the synthesized data set.
Here we see that the simulations using the prime model began with an $F1$ score of \textit{ca.} 0.6 that increased to 1 over time.
On the other hand, simulations using the null model began with an $F1$ score closer to 0 or 0.2 before gradually increasing to 1.
This shows that more accurate surrogate models for adsorption energies led to more accurate multiscale models, even initially.
Note also that the rate at which the $F1$ score improved was better when using \gls{MMS} than when using random sampling, especially when using the null model.
These data may suggest that the rate of improvement is governed by the acquisition strategy while the initial performance is governed by the model.

Figure~\ref{fig:syn_results}b shows how the \gls{dF} changes at each point in the simulation of the synthesized data set.
The simulations using random search generally yielded higher \gls{dF} values.
This indicates slower convergence, which is consistent with the slower $F1$ increase seen in the random search curves Figure~\ref{fig:syn_results}a.
Note also how the \gls{dF} values for the \gls{MMS}-prime simulation decreased at around 500 batches, which is the number of batches it took the $F1$ score to reach \text{ca.} 1.
Lastly, we note that the \gls{dF} values for the \gls{MMS}-null simulation were often zero.
This is because the null model was a ``stiff'' learner that did not result in any multiscale modeling changes unless a low-coverage adsorption site was found.
This shows that slow-learning models may result in relatively low \gls{dF} values, which may necessitate higher $\kappa$ values to offset this behavior.
In other words:  worse models may need longer horizons before stopping the discovery to mitigate the chances of missing important information.

These simulations provided us with an estimate of the improvement in active classification that we may get from using \gls{MMS}.
With the synthesized data set, we saw that the \gls{MMS}-with-null case achieved an $F1$ score of $\sim$0.6 after \textit{ca.} 250 batches (or 50,000 queries).
This was over seven times faster than the random-sample-with-null case, which achieved an $F1$ score of $\sim$0.6 after \textit{ca.} 1,800 batches (or 360,000 queries).
When using the prime model, \gls{MMS} was able to achieve an $F1$ score of $\sim$0.75 in 200 batches, while the random search achieved this same performance in \text{ca.} 1,200 batches, or six times slower.

\subsection{Performance on \gls{DFT} Data}

\begin{figure*}
    \centering
    \includegraphics[width=\textwidth]{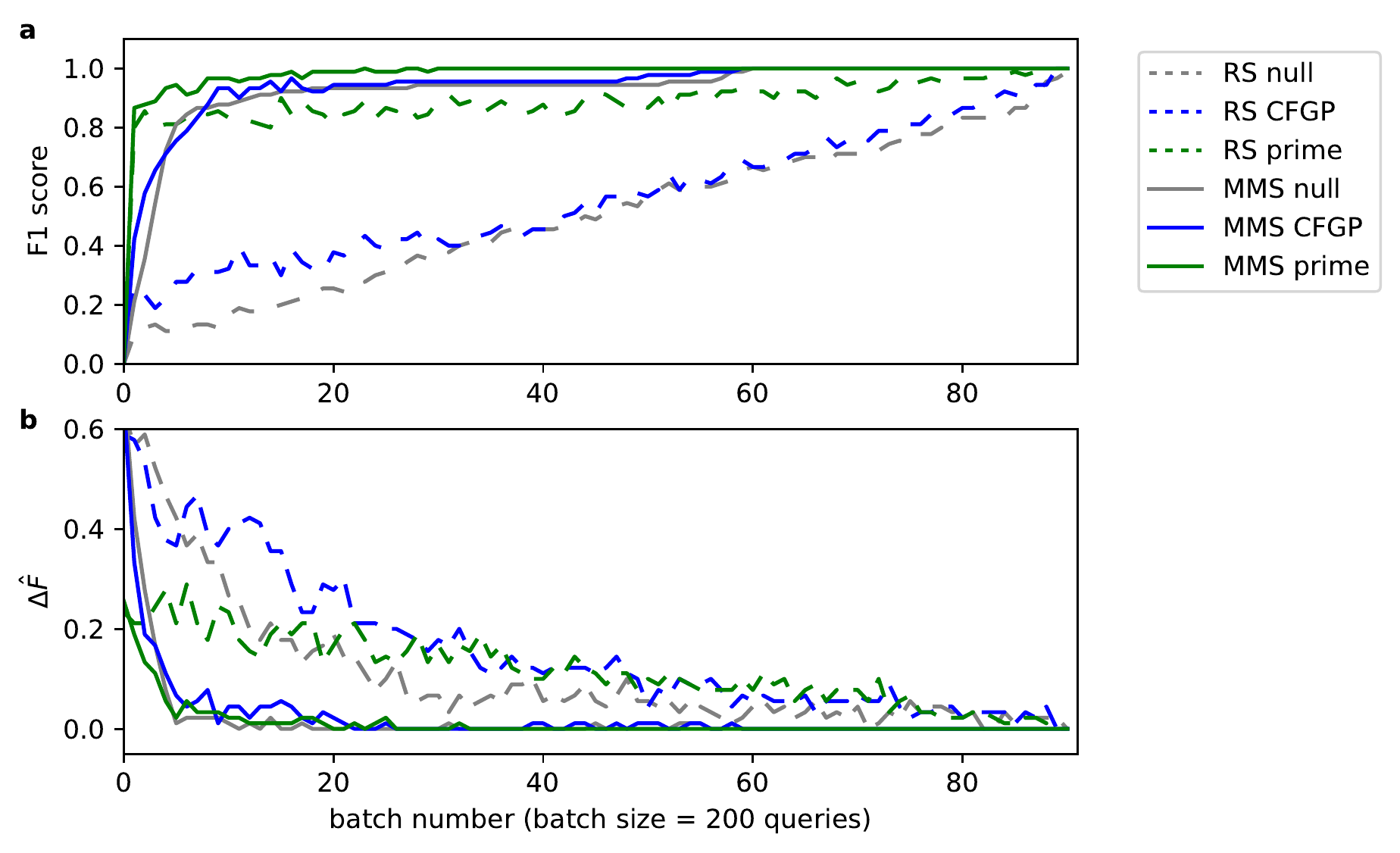}
    \caption{Performance and convergence results for the simulations on the \gls{GASpy} dataset.
        \textbf{a.} $F1$ score of the multiscale model during simulation of the \gls{GASpy} dataset.
        \textbf{b.} \gls{dF} of the multiscale model during simulation of the synthesized data.
        RS represents ``random search'' while \gls{MMS} represents \glsentrylong{MMS}.
        }\label{fig:gasdb_results}
\end{figure*}

Figure~\ref{fig:gasdb_results} shows the $F1$ score and the \gls{dF} of the multiscale model at each point in the simulation of the \gls{GASpy} data set.
Interestingly, the system performance when using \gls{CFGP} was similar to the performance when using the null model, both of which were overshadowed by the relatively good performance when using the prime model.
This suggests that there is a large room for improvement for the \gls{CFGP} model.
Note also how the \gls{MMS} strategy outperforms random sampling for this data set as well.

These simulations provided us with a second estimate of the improvement in active classification that we may get from using \gls{MMS}.
With the \gls{GASpy} data set, we saw that the \gls{MMS}-with-null case achieved an $F1$ score of $\sim$0.8 after \textit{ca.} 6 batches (or 1,200 queries).
This was over sixteen times faster than the random-sample-with-null case, which achieved an $F1$ score of $\sim$0.6 after \textit{ca.} 80 batches (or 16,000 queries).
When using the prime model, both \gls{MMS} and random search were able to achieve an $F1$ score of $\sim$0.8 after only a single batch.

\subsection{Recommended diagnostics}

\begin{figure*}
    \centering
    \includegraphics[width=\textwidth]{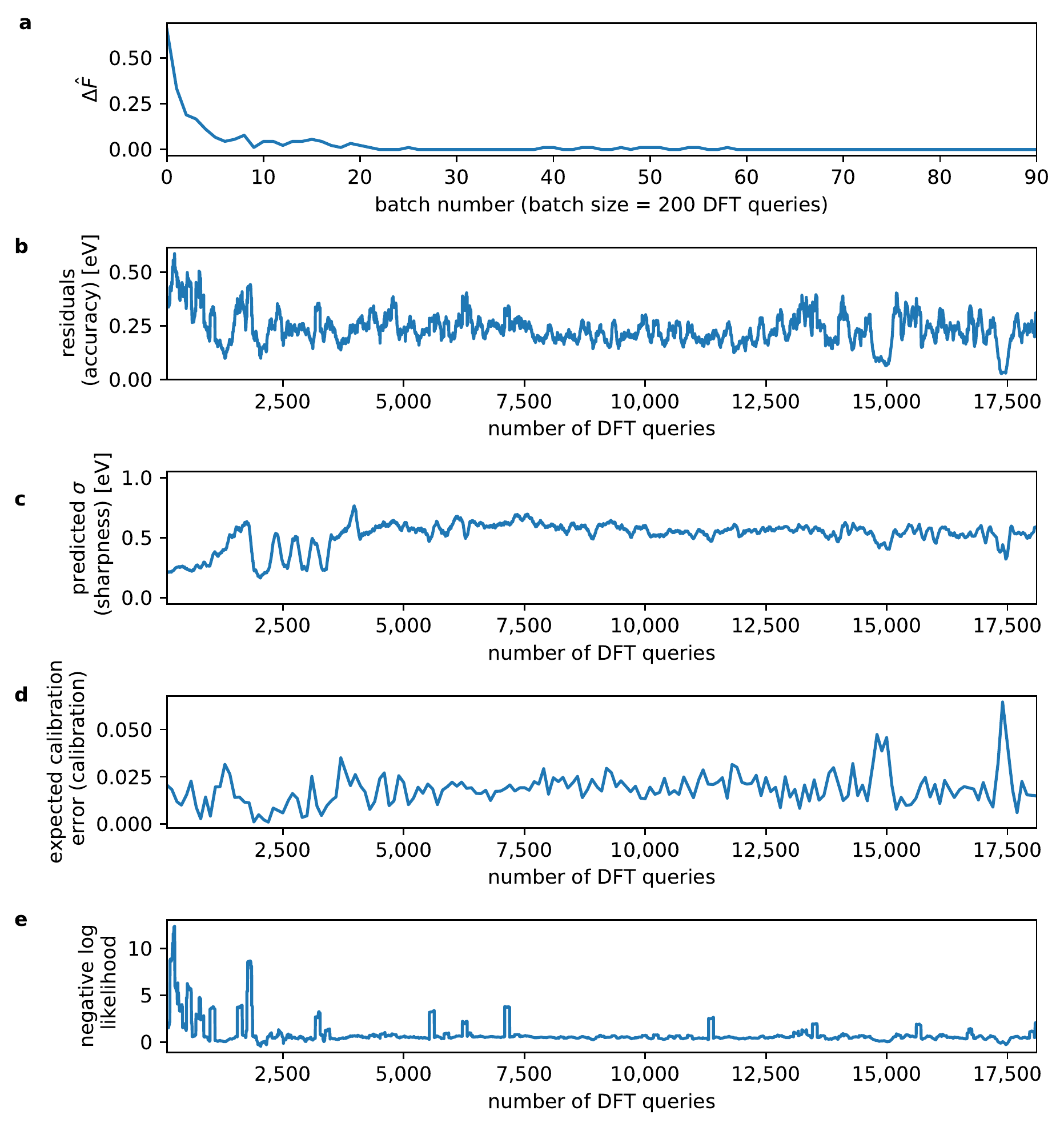}
    \caption{Example of diagnostic plots that we recommend monitoring during an active discovery campaign:
        \textbf{a.} \glsentryfull{dF};
        \textbf{b.} residuals between the real data and the surrogate model's predictions;
        \textbf{c.} expected calibration error\cite{Tran2020} of the surrogate model;
        \textbf{d.} the predicted uncertainties of surrogate model in the form of the predicted standard deviation ($\sigma$); and
        \textbf{e.} the negative-log-likelihood of the surrogate model.\cite{Tran2020}
        These results were simulated by using the \glsentryfull{MMS} method with the \glsentryfull{CFGP} model on the \gls{GASpy} dataset.
        For clarity of visualization, we plotted rolling averages of all values in this figure using a window of 100 queries (excluding the \gls{dF} values, where no averaging was done)}\label{fig:diagnostics}
\end{figure*}

We note that the $F1$ scores illustrated in Figures~\ref{fig:syn_results}a and~\ref{fig:gasdb_results}a cannot be calculated without knowing all the true classes, which is not possible to know during a real discovery process.
We need metrics to monitor the behavior of both our discovery algorithm.
We recommend monitoring the \gls{dF} as well as the accuracy, calibration, and sharpness (i.e., the magnitude of the predicted uncertainties) of the surrogate model over time.
Figure~\ref{fig:diagnostics} shows an example of such diagnostic metrics over the course our simulation that used \gls{MMS} and \gls{CFGP} on the \gls{GASpy} dataset.

\gls{dF} estimates the amount of overal improvement in the discovery process.
Sustained low values of \gls{dF} are a necessary but not sufficient indicator of convergence.
To improve our confidence in the predictive strength of \gls{dF}, we can test one of its underlying assumptions:  that the multiscale model becomes progressively more accurate as it receives more data.
This assumption is true when we replace surrogate model predictions with incoming \gls{DFT} results, but it is not necessarily true for unqueried points.
We can estimate the accuracy on unqueried points by calculating the residuals between the surrogate model and the incoming \gls{DFT} results (Figure~\ref{fig:diagnostics}b).
As each ``batch'' of queries is recieved, we compare the queried, true adsorption energies with the energies predicted by the surrogate model just before retraining---i.e., the predictions used to choose that batch.
Any improvements in accuracy on these points show that the overall, multiscale model is improving over time and that the \gls{dF} metric is an honest indicator of convergence.
Figure~\ref{fig:diagnostics}b shows that model accuracy improves within the first \textit{ca.} 10 batches (or 2,000 adsorption energy queries), but plateaus afterwards.
This indicates that, after 10 batches, improvements in overall classification accuracy came from receipt of additional \gls{DFT} data rather than improvements in surrogate model predictions.

Prediction accuracy of adsorption energies is not the only indicator of improved model performance.
If a surrogate model's accuracy does not change but its uncertainty predictions decrease/improve, then our confidence in the overall material classification may still improve.
Of course, improvements in uncertainty must not be obtained at the expense of worse calibration.
In other words, reductions in predicted uncertainties may also indicate improved model performance and better confidence in \gls{dF}, but only if the expected calibration error\cite{Tran2020} does not increase.
In our illustrative example, Figure~\ref{fig:diagnostics}c shows the predicted uncertainty while Figure~\ref{fig:diagnostics}d shows the calibration.
Unfortunately, the uncertainty predictions do not decrease over the course of the discovery process.
Note that all uncertainty and calibration estimates for each batch should be calculated using the surrogate model predictions used to choose that batch, just as was done for the residuals.

Lastly, we also recommend monitoring the negative-log-likelihood\cite{Tran2020} of the surrogate model for each incoming batch.
This metric incorporates model accuracy, calibration, and sharpness into a single metric.
Lower values of negative-log-likelihood indicate better model performance.
Figure~\ref{fig:diagnostics}e shows that this metric improves until \textit{ca.} 2,000 queries, after which it stagnates.
This is consistent with the improvement in accuracy until 2,000 queries and subsequent stagnation of all performance metrics thereafter.


\section{Conclusions}

Here we created a multi-scale modeling method for combining atomic-scale \gls{DFT} results with surrogate/\gls{ML} models to create actionable plans for experimentalists---i.e., a classification of catalysts as ``worthy of experimental study'' or ``not worthy''.
We then coupled this modeling method with a \glsentryfull{MMS} strategy to perform automated catalyst discovery via active classification.
We tested this strategy on two hypothetical datasets using three different surrogate models, giving us an estimate on the range of performance we might see in the future.
In some cases, the results show up to a 16-fold reduction in the number of \gls{DFT} queries compared to random sampling.
The degree of speed-up depends on the quality of the \gls{ML} model used, the homogeneity of the search space, and the hyperparameters used to define convergence of the active classification.
Speed-up estimates on more realistic use cases show a more conservative 7-fold reduction in number of \gls{DFT} queries.
Lastly, we provide a set of recommended diagnostic metrics to use during active classification (Figure~\ref{fig:diagnostics}):  \gls{dF} and the \gls{ML} model's residuals, uncertainty estimates, and calibration.

Our results elucidated a number of qualitative behaviors of active classification.
First, we observed that higher-quality \gls{ML} models yielded better initial performance of the classification process.
Conversely, we observed that higher-quality sampling strategies yielded better rates of improvement over time.
We also observed that our latest \gls{ML} model (\gls{CFGP}) yielded performance closer to a naive, ignorant model than to a perfect, omniscient model.
This suggests that there is a relatively large amount of potential improvement left in the \gls{ML} modeling space.
Next, we observed that better sampling strategies (as quantified by $F1$ score) led to lower rates of change in classes (as quantified by \gls{dF}), suggesting that \gls{dF} may be an indicator of sampling strategy performance.
Conversely, we observed that slow-learning \gls{ML} models may also reduce \gls{dF}.
This phenomena could be counteracted by using more conservative convergence criteria.
All these details were observed in specific and synthetic use cases though.
The behaviors seen here may not be observed in situations where search spaces and/or \gls{ML} models differ.

We encourage readers to focus on the main goals of this work:  (1) converting atomic-scale simulations and \gls{ML} models into actionable decisions for experimentalists, and (2) relaxing the active discovery process from an optimization/regression problem to a classification problem.
The ability to convert computational results into experimental recommendations helps us serve the research community better.
Simultaneously, relaxing the discovery process to a classification problem helps us prioritize exploration rather than exploitation, which is more appropriate for early-stage discovery projects.

We also recognize several future directions that may stem from this research.
Future work might include incorporation of \gls{DFT}-calculated surface stability by performing weighted averaging of surface activities when calculating bulk activities.
Future work may also include cost-weighted sampling such that less computationally intensive calculations are chosen more frequently than more intensive ones, which may improve discovery rates in real-time.
Perhaps most importantly, future work should incorporate some ability to feed experimental data and information to computational sampling strategies---e.g., multi-fidelity modeling.


\section*{Data Availability}
The data that support the findings of this study are openly available in https://github.com/ulissigroup/catalyst-acquisitions/.

\begin{acknowledgments}
    This research used resources of the National Energy Research Scientific Computing Center, a DOE Office of Science User Facility supported by the Office of Science of the U.S. Department of Energy under Contract No. DE-AC02-05CH11231.  
    KGB acknowledges the U.S. Department of Energy, Office of Science, Office of Basic Energy Sciences, Data Science for Knowledge Discovery for Chemical and Materials Research program, under Award DESC0020392.
    WN was supported by U.S. Department of Energy Office of Science under Contract No. DE-AC02-76SF00515.  
\end{acknowledgments}

\bibliography{my_bib.bib}

\end{document}